\documentclass[journal=jctcce,manuscript=article]{achemso}

\usepackage{chemformula} 
\usepackage[T1]{fontenc} 

\definecolor{codegray}{gray}{0.9}

\author{Jitai Yang}  
\affiliation[jlu]
{Institute of Theoretical Chemistry, College of Chemistry, Jilin University, 2519 Jiefang Road,
	Changchun 130023, P.R.China}

\author{Yang Cong}  
\affiliation[jlu]
{Institute of Theoretical Chemistry, College of Chemistry, Jilin University, 2519 Jiefang Road,
	Changchun 130023, P.R.China}

\author{You Li}  
\affiliation[jlu]
{Institute of Theoretical Chemistry, College of Chemistry, Jilin University, 2519 Jiefang Road,
	Changchun 130023, P.R.China}

\author{Hui Li}
\affiliation{Institute of Theoretical Chemistry, College of Chemistry, Jilin University, 2519 Jiefang Road,
	Changchun 130023, P.R.China}
\email{ Prof_huili@jlu.edu.cn}

\title[An \textsf{jctcce} demo]
{A Machine Learning Approach Based on Range Corrected Deep Potential Model for Efficient Vibrational Frequency Computation}

\abbreviations{IR,NMR,UV}

\begin{document}
	
	\begin{tocentry}
		\includegraphics{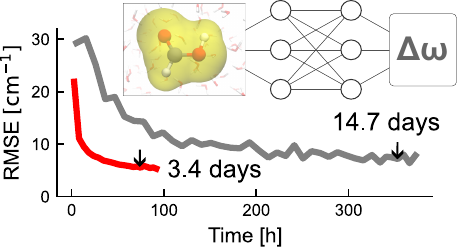}	
	\end{tocentry}
	
	\begin{abstract}
As an ensemble average result, vibrational spectrum simulation can be time-consuming with high accuracy methods. 
We present a machine learning approach based on the range-corrected deep potential (DPRc) model to improve computing efficiency. DPRc method divides the system into ``probe region'' and ``solvent region''; ``solvent-solvent'' interactions are not counted in the neural network.	
We applied the approach to two systems: formic acid \ch{C=O} stretching and MeCN \ch{C+N} stretching vibrational frequency shifts in water.
All data sets were prepared using Quantum Vibration Perturbation (QVP) approach.
Effects of different region divisions, one-body correction, cut-range, and training data size were tested.
The model with a single molecule ``probe region'' showed stable accuracy; it ran roughly ten times faster than regular DP and reduced the training time by about four.
The approach is efficient, easy to apply, and extendable to calculating various spectra.
	\end{abstract}
	
\section{Introduction}
Vibrational spectroscopy is a powerful experimental technique applied in various systems, including molecular clusters, solids, solutions, proteins, and surface systems.\cite{VS01,SFG01, VSprotein01, VS02, VS04, VS05, VS06} 
Theoretical simulations can help interpret the experimental spectra and gain additional insights, such as dynamic spectral diffusion, vibrational quantum effects, and the complexity of the local environment at the atomic level.
Computing ensemble average is necessary and the most time cost step if explicitly considering the dynamic and local chemical environment around the chromophore in simulation.
The average computing can base on a single chromophore with its local environment or all molecules in the system. 
When starting with a single chromophore molecule, there are fewer atoms, allowing for higher precision,  more rigorous treatment, and more analysis from a molecular view, such as solvatochromism and hydrogen bond analysis. 
However, introducing high precision or rigorous methods limits computing efficiency, and approximations must be reintroduced. \cite{Map1, Map2, Map3, Map4, Map5, Map6}

 Quantum Vibration Perturbation (QVP) can handle molecular quantum vibrational effects in spectrum simulation.\cite{Cong2022, Zhao2022QVP2, Yin2019QVP4, Olson2017QVP5, Xue2017QVP3} With contracted and localized basis from potential optimized discrete variable representation (PODVR)\cite{PODVR1992}, QVP approach is affordable in picosecond time scale or calculating tens of thousands of frequencies but still challenging in nanosecond time scale. Multi-dimension vibrational modes coupling problem also poses rigorous efficiency requirements.

Machine learning has grown rapidly in recent years and can help many efficiency problems in computational chemistry. 
We are interested in vibrational frequency calculation with machine learning in this work. There have been many related studies.\cite{VSMLCho2020, VSMLJaingJun2021, VSMLJiangBin2020, VSMLJiangJun, VSMLMullerSCM, VSMLSkinner, Add0Jiang2021,Add2Yao2018,  Add1Gandolfi2020}

Skinner group used artificial neural networks (ANN) with atom-centered symmetry functions (ACSFs) to reduce their map method's errors. Their final errors were higher than in later works.
\citeauthor{VSMLCho2020} used feed-forward neural network (FFNN) and convolutional neural network (CNN) with ACSF and polynomial functions descriptors. They used a small training dataset (1500 training data points). CNN performs slightly better than FFNN does in their work.
However, their training curve showed that CNN ran into overfitting quickly. The small data size possibly caused this. FFNN models showed a steady and slow decline during training.\cite{VSMLCho2020}
Cho group then applied FFNN to the OH stretch vibration of water with a same size dataset and obtained results with similar accuracy.\cite{VSMLChowater2021}
Jiang Jun group applied multi-layer perceptron (MLP) and Coulomb matrix (CM) to calculate neighboring couplings and frequencies of amide I in the protein system, achieving a high accuracy of less than several wavenumbers with harmonic approximation.\cite{VSMLJiangJun}
	
Considering a chromophore in its local chemical environment, a natural idea is to focus on the chromophore-chromophore atoms and chromophore-environment interactions, followed by environment-environment interactions. Additionally, for nonbonded interactions between neutral molecules, the effective distance is limited and is generally considered to be about 10~\AA. Many works have tried to treat the interactions of different atoms separately, and their interaction cut-off radii are usually around 6~\AA.\cite{VSMLCho2020, VSMLSkinner, VSMLJiangJun}
This picture is similar to the quantum mechanical/molecular mechanical (QM/MM) method,\cite{QMMMGao, QMMMQiang} and the deep potential range-corrected (DPRc) method is a machine learning method developed for computing the QM/MM energy correction, which able to treat the interactions separately.
As far as we know, there has yet to be any work on spectrum simulation utilizing the DPRc model. 

This work presents a machine learning approach based on DPRc to map structure to instantaneous vibrational frequency shift. We tested different interaction region divisions and 6/10~\AA\ cut ranges. All data sets were prepared with the QVP method. We tested the deep potential (DP) model as a limit case of DPRc for comparison.
	\section{Method}
	\subsection{Quantum vibration perturbation (QVP)}
	The semi-classical equation is common in computing the infrared (IR) spectral lineshape $I(\omega)$\cite{Skinner2012WT}
	\begin{equation}
		I(\omega) \propto \Re \int_0^{\infty} \mathrm{d} t e^{i \omega t}\left\langle\mathbf{m}(t) \mathbf{m}(0) e^{-i \int_0^t d \tau\left\{\omega(\tau)-\frac{i}{2 T_1(\tau)}\right\}}\right\rangle \label{eqn:semiCLS}
	\end{equation}
where $\boldsymbol{\mathrm{m}}$ is the vibrational transition dipole moment, $T_1$ is the relaxation time of the vibration, and the large angular brackets denote an ensemble average, 
$\omega(\tau)$ is
instantaneous vibrational frequency. 
The relaxation time $T_1$ is usually from an experiment result. With Condon approximation, we assume that the vibrational transition dipole moment $\boldsymbol{\mathrm{m}}$ is a constant. And we assume the orientation of $\boldsymbol{\mathrm{m}}$ is consistent with the vibrational local mode vector. Calculating the instantaneous frequency $\omega(\tau)$ costs time and contains effects like vibrational coupling and solvatochromism.
QVP approach treats $\omega(\tau)$ by combining molecular quantum vibration and molecular dynamics with perturbation theory.

 At first, we build reference quantum vibrational states. The Cartesian coordinates of the chromophore are expanded in terms of the normal mode vectors $\boldsymbol{\xi}_i(i=1, 2, ..., N-6)$ in equilibrium geometry $\boldsymbol{\mathrm{R}}_\mathrm{e}$,
 \begin{equation}
 	\mathbf{R}^{\mathrm{ch}}=\mathbf{R}_{\mathrm{e}}+\sum_i Q_i \boldsymbol{\xi}_i \label{eqn:NMexpand}
 \end{equation}
where $\boldsymbol{\mathrm{R}}^\mathrm{ch}$ is the Cartesian coordinates matrix of a nonlinear chromophore, and $Q_i(i=1, 2, ..., N-6)$ is the $i$th vibrational coordinate. 

We usually have one vibration we care most about to be treated with perturbation theory, like C=O stretch in formic acid, and mark the vibrational quantities corresponding to the vibration with subscripts ``s''.
The potential energy as a function of the mode $\boldsymbol{\xi}_\mathrm{s}$ is
\begin{equation}
	\hat{V}_0=V_0\left(Q_{\mathrm{s}}\right)=E_{\mathrm{BO}}\left(\mathbf{R}_{\mathrm{e}}+Q_{\mathrm{s}} \boldsymbol{\xi}_{\mathrm{s}}\right)
\end{equation}
$Q_\mathrm{s}$ is the vibrational coordinate of ``s'', and $E_\mathrm{BO}$ indicates the Born-Oppenheimer potential energy determined by an \latin{ab initio} method. The one-dimension vibration Schr\"odinger equation is solved with DVR\cite{DVR} and PODVR\cite{PODVR1992}. As a result, we choose an isolated chromophore as the reference states.
Next, we ``embed'' the vibrational degrees of freedom into the sample frames of molecular dynamics (MD) trajectory, where the dynamic chemical environment is added. More details about ``embed'' are in our previous work\cite{Cong2022}. Then instantaneous vibrational frequencies are obtained by Rayleigh-Schr\"odinger Perturbation Theory (RSPT). The perturbation potential $\hat{V}^{\prime}$ is
\begin{equation}
	\hat{V}^{\prime}=V^{\prime}\left(t ; Q_{\mathrm{s}}\right)=E_{\mathrm{BO}}\left(\mathbf{R}\left(t ; Q_{\mathrm{s}}\right)\right)-E_{\mathrm{BO}}\left(\mathbf{R}_{\mathrm{e}}+Q_{\mathrm{s}} \boldsymbol{\xi}_{\mathrm{s}}\right) \label{eqn:pert}
\end{equation}

The transition frequency $\omega^1$ between the first excited and ground states at first-order perturbation (QVP1) is 
\begin{equation}
	\begin{aligned}
		 \hbar\omega^1(t) & =E_1^0-E_0^0+\left\langle\phi_1\left|\hat{V}^{\prime}\right| \phi_1\right\rangle-\left\langle\phi_0\left|\hat{V}^{\prime}\right| \phi_0\right\rangle \\
		& =\Delta E^0+\int \mathrm{d} Q_{\mathrm{s}} \Delta \rho_{\mathrm{s}}\left(Q_{\mathrm{s}}\right) V^{\prime}\left(t ; Q_{\mathrm{s}}\right)\\
		& =\Delta E^0+\sum ^{N^\mathrm{PO}}_{i=1} \mathrm{d} Q_{\mathrm{s},i} \Delta \rho_{\mathrm{s}}\left(Q_{\mathrm{s},i}\right) V^{\prime}\left(t ; Q_{\mathrm{s},i}\right)		
	\end{aligned} \label{eqn:deltaE1}
\end{equation}
$ | \phi_0 \rangle$, $| \phi_1 \rangle$ are unperturbed wavefunctions of the ground state and the first excited state, 
$\Delta\rho_{\mathrm{s}}$ is the density difference between the first excited and ground state. The convergence of the perturbation orders should be tested.

In the last line of the eqn~\ref{eqn:deltaE1}, we use
PODVR basis functions/points from isolated chromophore to treat the integral. PODVR needs a few localized basis functions/points to converge, which enhances efficiency. Using economic computation approaches in eqn~\ref{eqn:pert}, such as DFT or a semi-empirical method, will also save time while losing some accuracy.
Finally, the lineshape can be calculated by instantaneous time-dependent vibrational frequency shifts $\Delta\omega$ from $\omega$ of reference states. More details about QVP are in our previous work.\cite{Cong2022, Xue2017QVP3}

\subsection{Deep potential range correction (DPRc) for frequency shift computing}


DPRc is a derivative version of the Deep Potential (DP) model. This section introduces DP first, followed by DPRc. DP and DPRc usually treat energies and sometimes treat electric dipole moments. We restate the theory here in frequency shift calculation context for combination with QVP and spectrum simulation.

In the DP model, the final output is a sum of atomic contributions. Instead of energy, the physical quantity here is the vibrational frequency shift $S$,
	\begin{equation}
		S=\sum_{i=1}^N S_i \label{eqn:fsum}
	\end{equation}
	
	\begin{equation}
		S_i=\mathcal{N}(D(\widetilde{\mathcal{R}})) \label{eqn:net}
	\end{equation}
	
The atomic contribution $S_i$ is a neural network of hidden layers $\mathcal{N}$. Layer numbers and form of $\mathcal{N}$ are hyperparameters. 
$D(\widetilde{\mathcal{R}})$ is the input layer and the ``descriptor'' array, which contains a local embedding network to reduce the dimensions in $\widetilde{\mathcal{R}}$.\cite{Zhang2018}. 
	
	$\widetilde{\mathcal{R}}$ is the ``environment matrix'':
	\begin{equation}
		(\widetilde{\mathcal{R}}_i)_{j a}= \begin{cases}s(R_{i j}), & \text { if } a=1 \\ s(R_{i j}) X_{i j} / R_{i j}, & \text { if } a=2 \\ s(R_{i j}) Y_{i j} / R_{i j}, & \text { if } a=3 \\ s(R_{i j}) Z_{i j} / R_{i j}, & \text { if } a=4\end{cases} \label{eqn:R}
	\end{equation}
	$i$ is the $i$th center atom, $j$ is the $j$th atom around $i$,  $R_{ij}$, $X_{ij}$, $Y_{ij}$ and $Z_{ij}$ are defined as \textit{relative} coordinates with  $R_{ij} = R_j - R_i$ and $R_{ij} = (X_{ij}, Y_{ij}, Z_{ij})$, $s\left(R_{i j}\right)$ is a switching reciprocal distance function that controls the range of the environment to be described. 
	
	$s(R_{i j})$ is
	\begin{equation}
		s(R_{i j})= \begin{cases}
			\frac{1}{R_{i j}},                                                                                 & \text { if } R_{i j} \leq R_{\mathrm{on}}             \\
			\frac{1}{R_{i j}}
			\left\{
			\left(
			\frac{R_{i j}-R_{\text {on }}}{R_{\text {off }}-R_{\text {on }}}
			\right)^3 
			\left(
			-6\left(
			\frac{R_{i j}-R_{\text {on }}}{R_{\text {off }}-R_{\text {on }}} 
			\right)^2
			+15 \frac{R_{i j}-R_{\text {on }}}{R_{\text {off }}-R_{\text {on }}}  -10 
			\right) + 1
			\right\},   &   \text { if } R_{\mathrm{on}}<R_{i j}<R_{\mathrm{off}}\\
			0,                                                                                                 & \text { if } R_{i j} \geq R_{\text {off }}
		\end{cases} \label{eqn:switchF}
	\end{equation}
	The smooth function in the second line of the equation is for numerical stability. If a neighboring atom is within a distance of $R_\mathrm{on}$, the atom will have full weight. The weight smoothly decreases between $R_\mathrm{on}$ and $R_\mathrm{off}$.
	
Deep Potential Range Correction (DPRc) was developed to correct the energy of the QM/MM approach. DPRc corrects energy associated with QM-QM and QM-MM interactions. Additionally, it should not alter interactions between MM atoms.
 Similarly, we expect the model to focus on chromophore-chromophore and chromophore-surrounding interactions in the frequency shift calculations. In order to meet the demand, we divide the system into "probe region" and "solvent region" and treat region-region interactions with the switching function:
	\begin{equation}
		s_{ij}(R_{ij})=
		\begin{cases}
			0,                                                                                                                                                                                                                                                                                                                   & \text{ if } i j \in \text{Solvent}                           \\
			\frac{1}{R_{i j}},                                                                                                                                                                                                                                                                                                   & \text { if } i j \in \text{Probe}                          \\
			\text{if}~i \in \text{Probe} \wedge j \in \text{Solvent} \text { or } i \in \text{Solvent} \wedge j \in\text{Probe} \text { : }                                                                                                                                                                                       &                                                            \\
			\frac{1}{R_{i j}},                                                                                                                                                                                                                                                                                                   & \text { if } R_{i j} \leq R_{\text {on }}     \\
			\frac{1}{R_{i j}}
			\left\{
			\left(
			\frac{R_{i j}-R_{\text {on }}}{R_{\text {off }}-R_{\text {on }}}
			\right)^3 
			\left(
			-6\left(
			\frac{R_{i j}-R_{\text {on }}}{R_{\text {off }}-R_{\text {on }}} 
			\right)^2
			+15 \frac{R_{i j}-R_{\text {on }}}{R_{\text {off }}-R_{\text {on }}}  -10 
			\right) + 1
			\right\}, & \text{ if }  R_{\text {on }}  < R_{i j} <  R_{\text {off}} \\
			0,                                                                               \                                                                                                                                                                                                                                   & \text{ if } R_{i j} \geq R_{\text {off}}
			
		\end{cases} \label{eqn:rangeCor}
	\end{equation}

	Furthermore, though it is unnecessary in the frequency shift calculation, we can keep the one-body contribution of solvent region atoms to zero:
	\begin{equation}
		S_i= \begin{cases}
			\mathcal{N}(D(\widetilde{\mathcal{R}})), & i \in \mathrm{Probe} \\
			\mathcal{N}(D(\widetilde{\mathcal{R}}))-S_i^{(0)},    & i \in \mathrm{Solvent}
		\end{cases} \label{eqn:onebody}
	\end{equation}
	$S_i^{(0)}$ is the one-body contribution to the frequency shift
	\begin{equation}
		S_i^{(0)}=\mathcal{N}(D({{0}}))
	\end{equation}

Also, one can cancel the one-body contribution of specific components in the system, like water,

	\begin{equation}
 S_i= \begin{cases}
		\mathcal{N}(D(\widetilde{\mathcal{R}})), & i \notin \mathrm{Water} \\
		\mathcal{N}(D(\widetilde{\mathcal{R}}))-S_i^{(0)},    & i \in \mathrm{Water}
	\end{cases} \label{eqn:onebody_wt}
\end{equation}

	With the switch function of eqn.~\ref{eqn:rangeCor}, the model excludes the interactions based on different regions and distances between atoms.
	
	If the probe region has all atoms without one-body correction, DPRc is equal to the regular DP model. If the probe region has only one atom, as in Figure~\ref{fgr:shm}b, then only distances between the center atom and other atoms will count. However, the surrounding-surrounding interactions remain as ``atom types'' in the neural network. The final network expression still corresponds to a many-body form rather than a two-body form. More details about DPRc are in the work of \citeauthor{Zeng2021}\cite{Zeng2021}
%
%
%

	\subsection{Computational details}


\textbf{Formic acid solution}: We used the MD trajectory from our previous spectrum simulation work\cite{Cong2022}. The dilute solution is consist of one formic acid and 471 water molecules in a simulation cubic periodic box of 25~\AA. We ran the MD with QM/MM approach. The AM1 semiempirical method\cite{am1} treated formic acid, and the TIP3P model\cite{tip3p} treated waters. All MD ran with the CHARMM package\cite{charmm}. The system equilibrated for 500~ps in constant NVT at 298.15~K first, after which a 50~ps simulation was carried out for QVP analysis using a step length of 1~fs. 

In QVP analysis, We calculated the isolated formic acid optimized structure, normal modes and reference quantum vibrational states in CCSD(T)-F12b/cc-pVTZ-F12 level\cite{F12-1, F12-2, basis-pvnz}. The calculation ran with the MOLPRO v2012.1 package\cite{MOLPRO, MOLPRO-WIREs, MOLPRO2020}. 49999 frames in the MD trajectory were calculated to get the perturbation energies in GFN2-xTB(Geometry, Frequency, Non-covalent, eXtended TB) level\cite{GFN2}. 49989 frames converged in energy calculation. The frequency shifts converged in first order Rayleigh-Schr\"odinger perturbation. The calculation ran with xTB v6.6.0 package\cite{GFN2}. The linear scaling method was not applied. After QVP analysis, the frames and correspond frequency shifts were reformatted for the DeePMD-kit program.

Total 49989 frames and correspond frequency shifts consisted the data sets for machine learning. Earlier 80\% in MD trajectory, 39992 frames and frequency shifts consisted the training data set. The left 20\%, 9997 frames and frequency shifts were chosen as validation and test data sets. 

\begin{figure}[H]
	\includegraphics{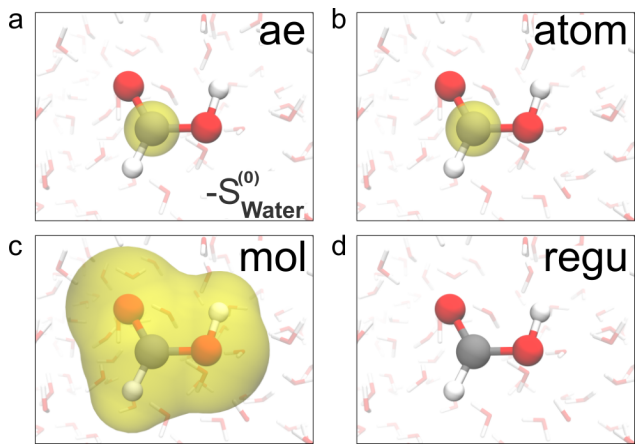}
	\caption{Schematic of models \textit{ae} (a), \textit{atom} (b), \textit{mol} (c), and \textit{regu} (d). The molecule in the center is formic acid surrounded by water molecules. Atoms in the yellow bubbles of \textit{ae, atom}, and \textit{mol} are selected as ``probe region''. The minus term in \textit{ae} means the isolated atomic contribution of water is subtracted (eq. \ref{eqn:onebody_wt}). \textit{regu} is the regular DP model.}
	
	\label{fgr:shm}
\end{figure}

We tested four kinds of models named \textit{ae}, \textit{atom}, \textit{mol}, and \textit{regu}. The schematic is placed in Figure \ref{fgr:shm}:
\textit{ae}, has the carbon atom of formic acid as probe region and canceled water one-body contribution as in eqn. \ref{eqn:onebody_wt}, the name is from the keyword in the DeepMD-kit program;
\textit{atom}, has the carbon atom of formic acid as probe region; 
\textit{mol}, has all formic acid atoms as probe region;
\textit{regu}, regular DP model, is a limit situation with all atoms in the probe region.

We used two cut range settings named \textit{s} and \textit{l} (short and long), corresponds to 6~\AA\, and 10~\AA\, cut-range ($R_\mathrm{off}$) respectively. $R_\mathrm{on}$ was set as 75\% of $R_\mathrm{off}$. Other hyperparameters and more details were in SI files. The machine learning ran with DeePMD-kit v2.0.3 CPU version\cite{Zhang2018} and used 8 CPUs for parallel computing. 

\textbf{MeCN solution}: The dilute solution consisted of one acetonitrile molecule and 721 water molecules in a simulation cubic periodic box. We implemented classical molecular dynamics simulations using TIP3P water model\cite{tip3p}. All atoms ran with charmm36 force fields\cite{charmm36}. Bonds with hydrogen atoms were constrained. The system was minimized in 500 steps first. We ran a 200~ps equilibration in constant NVT at 300~K, followed by an 800~ps equilibration in constant NPT at 300~K and 1~atm to adjust the box size, then ran another 200~ps equilibration in constant NVT at 300~K. And 2~ns production run in constant NVT at 300K. The production ran with 1~fs per step and saved the frame every 40~fs. All MD ran with the GROMACS v2021.3 package\cite{GROMACS}. 50001 frames were saved. 

In QVP analysis, We calculated the isolated acetonitrile optimized structure, normal modes, and reference quantum vibrational states in CCSD(T)-F12a/AVTZ level\cite{F12-1, F12-2, basis-avnz}. The calculation ran with the MOLPRO v2012.1 package\cite{MOLPRO, MOLPRO-WIREs, MOLPRO2020}. Our calculation omitted couplings between normal modes. 
For frames in the MD trajectory, we first made clusters with the MeCN and waters 12~\AA\ around the center of mass of MeCN. We used the MDAnalysis package\cite{mda1, mda2} for operating coordinates. Then we calculated the perturbation energies in GFN2-xTB(Geometry, Frequency, Non-covalent, eXtended TB) level\cite{GFN2}. All clusters converged in energy calculation. The frequency shifts converged in first-order Rayleigh-Schr\"odinger perturbation; we chose second order for more accurate results. The calculation ran with xTB v6.6.0 package\cite{GFN2}. The linear scaling method was not applied. After QVP analysis, the frames and corresponding frequency shifts were reformatted for the DeePMD-kit program.

50000 frames and corresponding frequency shifts comprised the data sets for machine learning. 40000 frames and frequency shifts comprised the training data set. The left 20\%, 10000 frames and frequency shifts were chosen as validation and test data sets. 

The machine learning approach was the same as in the formic acid solution. In \textit{ae} and \textit{atom} models, the carbon atom bonded with the nitrogen atom was selected as the probe region.

	\section{Results and discussion}
	
	\subsection{Formic acid solution}
	
	We tested four kinds of models named \textit{ae}, \textit{atom}, \textit{mol}, and \textit{regu} with different region divides and 6/10~\AA\ (\textit{l}/\textit{s}) cutoff ranges. We trained these models to map structures to \ch{C=O} stretch frequency shifts. The schematic is in Figure \ref{fgr:shm}.

\begin{figure}[H]
	\includegraphics{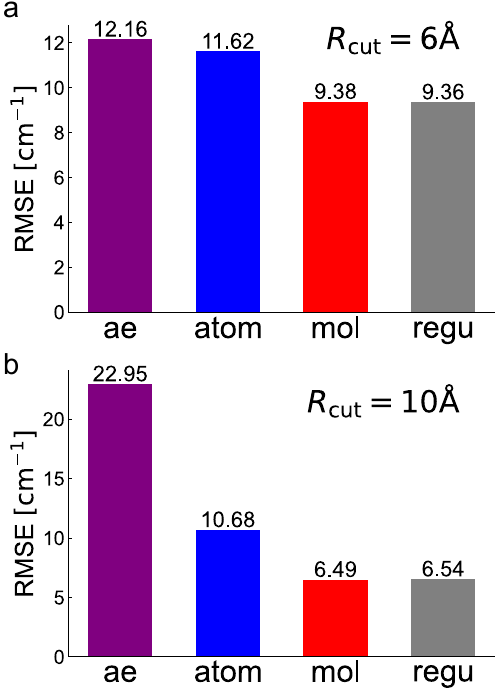}
	\caption{Root-mean-square errors in predicting frequency shifts of formic acid solution. (a), (b): \textit{ae, atom, mol,} and \textit{regu} results with 6/10~\AA\ cut-ranges.}
	\label{fgr:rmse2in1}
\end{figure}

RMSE results are in Figure \ref{fgr:rmse2in1}. \textit{mol}/\textit{l} and \textit{regu}/\textit{l} gave the best results. The worst result was from \textit{ae}/\textit{l}, which was roughly equal to predicting all shifts with the average data value. \textit{ae}/\textit{s} was worse than other models with short cut-range.

	The difference between \textit{ae} and \textit{atom} is \textit{ae} excludes single-body contributions of water. Our results indicate that the single-body correction is unsuitable for frequency shift prediction. This explains why \textit{ae} performed better with short cut-range because the model with short cut-range has less energy correction.
	
	%
The results for \textit{mol} and \textit{regu} are similar, indicating minimal effect from excluding environment-environment interactions. \textit{mol} showed slightly better results than \textit{atom}. \textit{mol} has more atoms in its ``probe region'' than \textit{atom}, allowing further solvent molecules to enter the net. So \textit{mol} contains more interactions. Long cut-range also yielded better results in all models. Another probable reason for \textit{mol} better than \textit{atom} is the completeness problem discussed by Jiang Bin group\cite{Zhang2021}; only one center atom in the probe region makes the problem worse.

\begin{figure}[H]
		\includegraphics{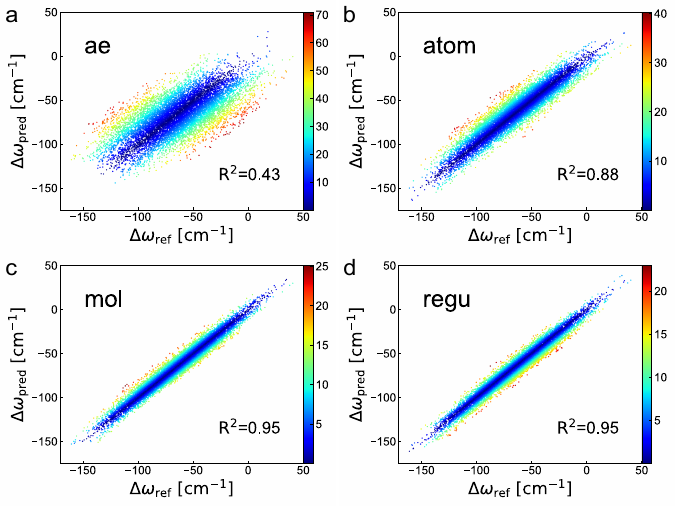}
		\caption{Error distribution in predicting frequency shifts of formic acid solution with \textit{ae} (a), \textit{atom} (b), \textit{mol} (c), and \textit{regu} (d) results with 10~\AA\ cut-range. The points' colors correspond to the absolute error values.}
		\label{fgr:distL4in1}
	\end{figure}

\begin{figure}[H]
		\includegraphics{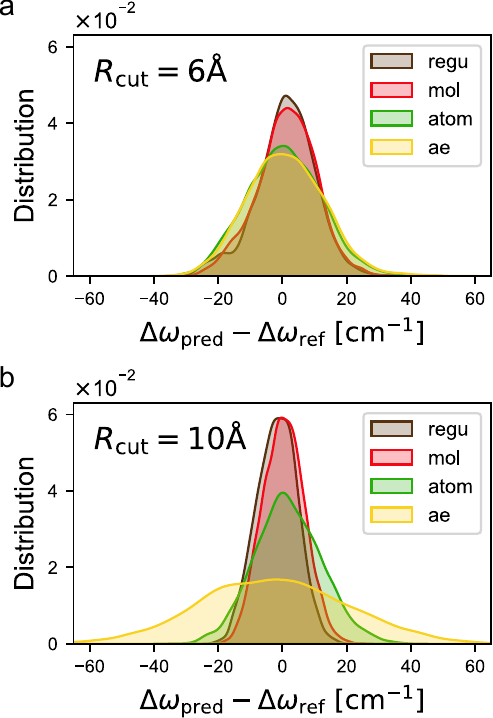}
		\caption{Kernel density estimate (KDE) plots of error distributions in predicting formic acid solution frequency shifts. (a), (b): \textit{ae, atom, mol,} and \textit{regu} results with 6/10~\AA\ cut-ranges.}
		\label{fgr:kde2in1}
	\end{figure}

The comparison of error distribution is shown in Figure \ref{fgr:distL4in1}, distribution of errors with short cut-range is in Figure S1. The Kernal Density Estimate (KDE) plot of error distribution is in Figure \ref{fgr:kde2in1}. The trend observed in the error distribution results is consistent with the RMSE results.


\begin{figure}[H]
		\includegraphics{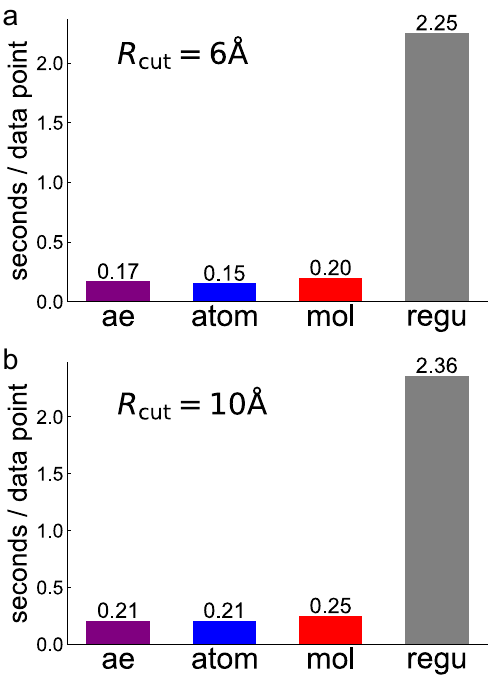}
		\caption{Running time of seconds per data point in predicting frequency shifts of formic acid solution. (a), (b): \textit{ae, atom, mol,} and \textit{regu} results with 6/10~\AA\ cut-ranges.}
		\label{fgr:runtime2in1}
	\end{figure}

	No significant difference in running time was found between \textit{ae}, \textit{atom}, and \textit{mol}, as shown in Figure \ref{fgr:runtime2in1}. The regular DP model \textit{regu} with the biggest network was the slowest.
In QVP or DVR methods, the most time cost step is computing energy on grid structures containing hundreds of environmental atoms. Semi-empirical methods like GFN2 can take one minute to get energy.
Five single-point energy calculations are necessary if we use five PODVR localized basis. Consequently, computing one frequency shift will take five minutes on one particular MD snapshot. So even the slowest model \textit{regu} is much faster than the usual method. 

\begin{figure}[H]
		\includegraphics{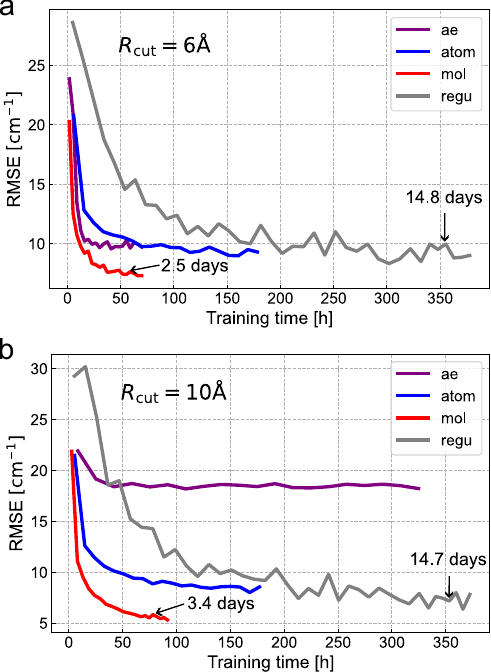}
		\caption{ Training curves in predicting frequency shifts of validation data of formic acid solution. (a), (b): \textit{ae, atom, mol, regu}
		 with 6/10~\AA\ cut-ranges. Results were smoothed for clarity.}
		\label{fgr:compare2in1}
	\end{figure}

However, as Figure \ref{fgr:compare2in1} shows, machine learning models need time to train. \textit{regu} took more than two weeks to get a stable training result, which limited its feasibility. In contrast, \textit{mol} models can obtain a stable result in 100 hours. We trained and tested all models in 8 CPUs. We believe running on GPU will be faster. As Figure S3c and S3d shows, we observed overfitting trends in \textit{mol} and \textit{regu} with 6~\AA\ cut range, but the RMSE results of validation are stable.

	
\subsection{MeCN solution}
	We tested three kinds of models named \textit{ae}, \textit{atom}, and \textit{mol} with different region divides and 6/10~\AA\ (\textit{l}/\textit{s}) cutoff ranges. \textit{regu} test was not performed due to time cost. We trained these models to map structures to \ch{C+N} stretch frequency shifts. 
		\begin{figure}[H]
		\includegraphics{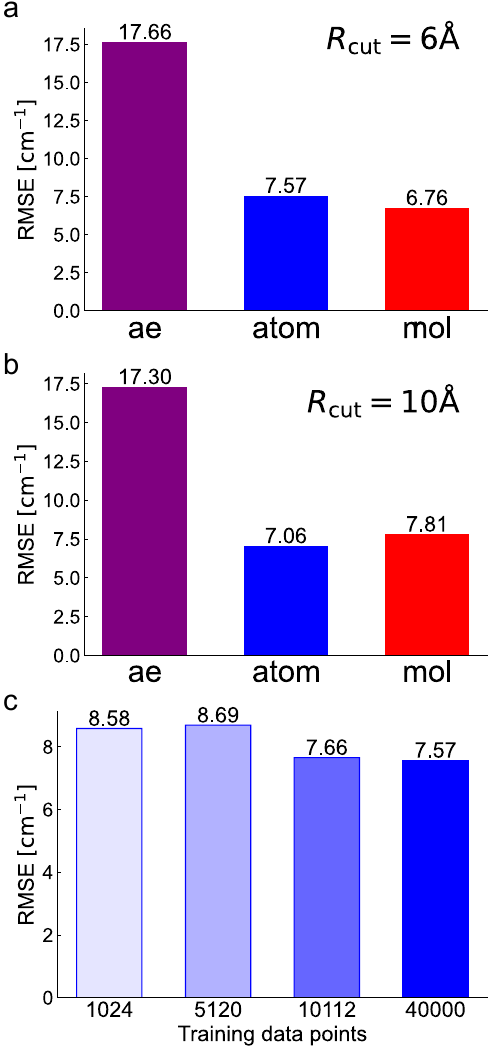}
		\caption{Root-mean-square errors in predicting frequency shifts of MeCN solution. (a), (b):  \textit{ae, atom, mol} results with 6/10~\AA\ cut-ranges. (c): \textit{atom}'s performance with different training data sizes with 6~\AA\ cut-range.}
		\label{fgr:acnRmse3in1}
	\end{figure}
	
	RMSE results are in Figure \ref{fgr:acnRmse3in1}a and Figure \ref{fgr:acnRmse3in1}b. Cho M group studied the
nitrile stretch mode of acetonitrile in water with differential evolution algorithm approach\cite{ChoMeCN}. Their system contained max 50 water molecules. Our best results have similar accuracy with about 260 water molecules.

\textit{ae} performed worst in three models, same as in the formic acid solution. 

MeCN solution results are insensitive when changing the cut range from 6~\AA\ to 10~\AA: Explicit difference between short and long cut ranges in formic acid solution was not observed. Figure \ref{fgr:acnRmse3in1}a and Figure \ref{fgr:acnRmse3in1}b also show little difference between \textit{atom} and \textit{mol}. 

Figure \ref{fgr:acnRmse3in1}c shows \textit{atom}'s performance with different data sizes with 6~\AA\ cut-range. We used about 40000 points to train formic acid and MeCN solution because it is the usual data number to get a converged spectrum. Results in Figure \ref{fgr:acnRmse3in1}c show we can use fewer data points when using DPRc. The accuracy decreased slightly as the amount of training data decreased. Our smallest training data size has 1024 data points, close to the data size used in the Cho group's work and their study on N-methylacetamide in water\cite{VSMLCho2020}.

We observed no explicit overfitting trend in results except in the training of \textit{atom} with 1024 training data points with 6~\AA\ cut-range. We let the training continue for days after we thought the RMSE in validation converged. The RMSE in validation was stable after the training curve showed an overfitting trend, as in Figure S13a.
More RMSE results and training curves are in supporting information files.
	

	\section{Conclusion}
	To improve the computational efficiency of the vibrational spectrum simulation, we present a machine learning method based on DPRc and QVP. We divided the system into ``probe region'' and ``solvent region''; ``solvent-solvent'' interactions were not counted in the neural network.
Effects of different divisions, ``one-body correction'', cut range, and training data size were tested.
We applied the models in two systems: formic acid \ch{C=O} stretching and MeCN \ch{C+N} stretching vibrational frequency shift in water.

We found that: Excluding ``solvent-solvent'' interactions affected little accuracy and was efficient in the calculation. \textit{mol}, has all the chromophore molecule atoms in ``probe region'', showed the most stable results with RMSE under 10~$\mathrm{cm^{-1}}$ and was much faster in running and training than DP; ``one-body'' correction was unsuitable for computing frequency shift;
The cut range showed different effects in different systems. Cut range longer than 6~\AA\ in descriptors can improve the performance in formic acid solution system;
We got a stable result on 1024 training data points. Training data size was only tested on one model and one system. We believe 5000 training data points are enough for most models and systems; No explicit overfitting trends in most trainings. 

Our model validates the design of focusing on chromophore and can assist future long-time spectral sampling and vibrational modes coupling research. The approach is feasible, easy to apply, and can be extended to calculate various spectra, such as Raman and sum frequency generation spectrum.

	\begin{acknowledgement}
		
		This research was sponsored by the 2020-JCJQ Project
		(GFJQ2126-007) and the National Natural Science Foundation of China (Grants 22073035, 21773081, and 21533003).
		
	\end{acknowledgement}
	
	\begin{suppinfo}
		
		Error distribution and training curves plots of models and input files of 8 models which contain all parameters.

		\begin{itemize}
			\item SI.pdf: 
			\subitem Formic acid solution results: Error distributions with 6~\AA\ cut-range; Training curves over time and training steps.
			\subitem MeCN solution results: Training curves over time and training steps; Training curves of training data size test.
			\subitem Descriptions of some hyperparameters.
			\item inputs.zip: Input files example of 8 models.
		\end{itemize}
		
	\end{suppinfo}
	
	\bibliography{manuscript}
	
\end{document}